\documentclass[prl,twocolumn,showpacs,superscriptaddress,amssymb]{revtex4}

\usepackage{psfrag,graphicx}
\usepackage{dcolumn}
\usepackage{bm}
\usepackage{amsfonts,amssymb,amsmath}        

\usepackage{xcolor}

\newcommand{\be}{\begin{equation}}
\newcommand{\ee}{\end{equation}}
\newcommand{\bq}{\begin{eqnarray}}
\newcommand{\eq}{\end{eqnarray}}

\newcommand{\rf}[1]{(\ref{#1})}
\newcommand{\SIGMA}{\mbox{\boldmath${\sigma}$}}

\newcommand{\p}{{\bf p}}
\newcommand{\rr}{{\bf r}}
\newcommand{\ket}[1]{\left |#1 \right\rangle}
\newcommand{\bra}[1]{\left \langle #1 \right |}
\begin{document}

\title{Abelian Chern-Simons-Maxwell theory from a tight-binding model of spinless fermions} 

\author{Giandomenico Palumbo}
\affiliation{Department of Physics, University of Pavia, via Bassi 6, 27100 Pavia, Italy} 
\affiliation{School of Physics and Astronomy, University of Leeds, Leeds, LS2 9JT, United Kingdom}
\author{Jiannis K. Pachos}
\affiliation{School of Physics and Astronomy, University of Leeds, Leeds, LS2 9JT, United Kingdom}

\date{\today}


\begin{abstract}

Abelian Chern-Simons-Maxwell theory can emerge from the bosonisation of the $2+1$-dimensional Thirring model that describes interacting Dirac fermions. Here we show how the Thirring model manifests itself in the low energy limit of a two-dimensional tight-binding model of spinless fermions. To establish that we employ a modification of Haldane's model, where the ``doubling" of fermions is rectified by adiabatic elimination. Subsequently, fermionic interactions are introduced that lead to the analytically tractable Thirring model. By local density measurements of the lattice fermions we can establish that for specific values of the couplings the model exhibits the confining $2+1$-dimensional QED phase or a topological ordered phase that corresponds to the Chern-Simons theory. The implementation of the model as well as the measurement protocol are accessible with current technology of cold atoms in optical lattices.

\end{abstract}

\maketitle

{\em Introduction:--} Chern-Simons theories are topological quantum field theories that can support anyonic particles with exotic mutual statistics \cite{Wilczek}. In high energy physics these theories are encountered in the context of quantum anomalies \cite{Bell,Adler} and in the study of gauge theories \cite{Jackiw}. In the context of condensed matter Chern-Simons theories emerge as effective theories for the description of the fractional quantum Hall liquids \cite{Zhang,Read}, of the surface states of three-dimensional topological insulators \cite{Qi} or of graphene coupled to external magnetic fields \cite{Fialkovsky}. Double Chern-Simons theories, called BF theories, have recently found application in graphene when it is decorated with a variety of gauge fields \cite{Giandomenico}. 

An important high energy physics example that supports the Chern-Simons-Maxwell theory is the $2+1$-dimensional massive Thirring model. This model describes massive interacting Dirac fermions \cite{Kondo}. It is well known that in $1+1$ dimensions, there exists an exact mapping between this massive model and the bosonic sine-Gordon model \cite{Coleman}. In the $2+1$-dimensional case the bosonisation gives rise to the Abelian Chern-Simons-Maxwell theory in the large fermion mass limit \cite{Fradkin,Deser1}. In this letter we establish a new connection between relativistic quantum field theory and condensed matter physics. In particular, we derive the $2+1$-dimensional Thirring model from a tight-binding model of fermions in the following way. It is well known that Dirac fermions can faithfully describe the low energy behaviour of fermions tunnelling on a honeycomb lattice. An undesirable doubling in these fermionic modes results from the lattice nature of the system \cite{Nielsen}. Haldane \cite{Haldane} decorated the honeycomb lattice with next-to-nearest neighbour tunnelling couplings in such a way that the two Dirac modes acquire inequivalent energy gaps. Here we employ the adiabatic elimination procedure to freeze the dynamics with respect to one of the Dirac modes. Subsequently, we introduce interactions between the lattice fermions and obtain the $2+1$-dimensional Thirring model. While the Haldane model gives rise to the integer quantum Hall effect, the interactions introduced in this letter are exactly designed to produce fractionalisation of charge. Hence, a Chern-Simons theory emerges with quasiparticle excitations that are Abelian anyons. This theory is accompanied by an additional Maxwell term that can be either made negligible or dominant by controlling the interactions between fermions. It is worth noting that, similar to the Haldane model, our  model breaks time reversal symmetry without a magnetic field. Finally, to demonstrate the topological order of the tight-binding model, we employ the stabilisation of its ground state against arbitrary Wilson loop operators. This can be shown just by performing local density measurements of the lattice fermions. 

\begin{figure}
\includegraphics[width=\linewidth]{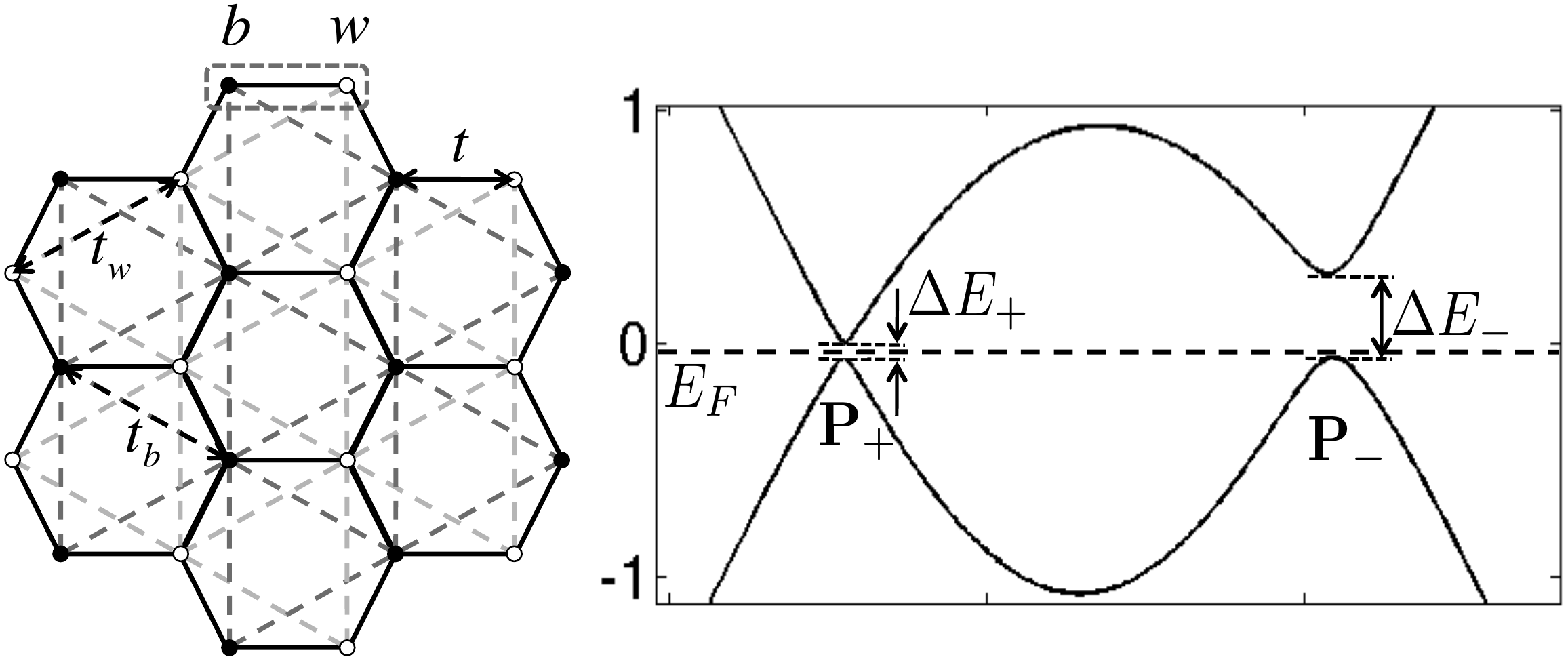}
\caption{\label{fig:EnergyGaps} {\em Left:} The honeycomb lattice with fermions tunnelling from one site to the neighbouring (coupling $t$) as well as to the next-to-neighbouring sites (couplings $t_b$ and $t_w$). The unit cell is depicted with two sites named $b$ and $w$. {\em Right:} The energy dispersion of the $t$- and $t_{b,w}$-terms of Hamiltonian \rf{Ham1} for momenta that cross trough both Fermi points, ${\bf P}_+$ and ${\bf P}_-$. The $t_{b,w}$-term opens up asymmetric energy gaps $\Delta E_+$ and $\Delta E_-$ to the corresponding Fermi points. We take the Fermi energy $E_F$ (depicted with dashed line) between both gaps so that only the lower band is completely filled. For $\Delta E_-\gg \Delta E_+$ we can adiabatically eliminate ${\bf P}_-$. Here we took $t=1$, $t_b=0.02$ and $t_w = 0.1$.}
\end{figure}

{\em From tight-binding to Thirring model:--} Let us start by describing the fermionic lattice with low energy behaviour given by the $2+1$-dimensional Thirring model. Consider the honeycomb lattice, shown in Fig. \ref{fig:EnergyGaps}, with fermionic modes placed at each lattice site. Fermions tunnel between nearest and next-to-nearest neighbouring sites. The unit cell of the lattice includes two sites that are named $b$ and $w$. The Hamiltonian of the system is given by
\bq
H =\!\!\!\!\!&& -t\sum_{\langle {\bf i},{\bf j}\rangle} (b_{\bf i}^\dagger w^{}_{\bf j}+w_{\bf i}^\dagger b^{}_{\bf j})-\sum_{\langle \langle {\bf i},{\bf j}\rangle\rangle}(-t_b b_{\bf i}^\dagger b^{}_{\bf j} +e^{i\phi_{{\bf ij}}} t_w w_{\bf i}^\dagger w^{}_{\bf j}) \nonumber \\
&&+
U\sum_{\bf i} b_{\bf i}^\dagger b^{}_{\bf i} w_{\bf i}^\dagger w^{}_{\bf i},
\label{Ham1}
\eq
where $t$ is the nearest neighbour tunnelling coupling and $t_b$ and $t_w$ are the next-to-nearest neighbour tunnelling couplings for the $b$ and the $w$ fermions, respectively. Finally, $U$ is the interaction coupling that is activated only between fermions of the same unit cell. For concreteness, we take all couplings, $t$, $t_b$, $t_w$ and $U$ to be real and positive. A complex phase factor $e^{i\phi_{\bf ij}}$ appears explicitly only in the next-to-nearest neighbour tunnelling term of the $w$-particles. Note also the minus phase factor in front of the $t_b$ couplings. 

For Fermi energies, $E_F$, close to half filling (see Fig. \ref{fig:EnergyGaps}) Hamiltonian (\ref{Ham1}) has the following characteristics. The low energy behaviour of the first $t$-term, $H_t$, is equivalent to graphene \cite{Novoselov,Semenoff}. The energy dispersion relation $E(\bf p)$ with respect to this term becomes zero for two isolated momenta, ${\bf P}_\pm = (0, \pm 4\pi/(3\sqrt{3}))$, called Fermi points. Expanding the Hamiltonian around these momenta gives
\be
H^{\pm}_t \approx -{3\over 2} t\int d^2{r} b^\dagger({\bf r})(\partial_x\pm i\partial_y)w({\bf r})+\text{H.c.},
\ee
where $b({\bf r})$ and $w({\bf r})$ are the continuous version of the fermionic operators and $\partial_{x,y}$ are partial derivatives in the two spatial dimensions. The Hamiltonians $H_t^\pm$ are gapless, so they describe massless Dirac fermions.

The second $t_{b,w}$-term, $H_{t_{b,w}}$, opens an energy gap at the Fermi points. We now take the phase acquired by $w$ fermions to be $\phi=-2 \pi/3$ for the direction $\bf n_{1}$=$(3/2, \sqrt{3}/2)$ ($-\phi$ for the direction $-\bf n_{1}$) and zero for the rest of the directions. Then, close to the two Fermi points, i.e. within the low energy approximation, $H_{t_{b,w}}$ assumes the following forms
\bq
H_{t_{b,w}}^{+}&\approx& -3\int d^2{r} t_{b}b(\rr)^\dagger b(\rr),\nonumber \\
H_{t_{b,w}}^{-}&\approx& 3\int d^2{r} \left[t_{w}w(\rr)^\dagger w(\rr)-t_{b}b(\rr)^\dagger b(\rr)\right].
\eq
These Hamiltonians give rise to the energy gaps $\Delta E_+ = 3 t_b$ for ${\bf P}_+$ and $\Delta E_- = 3(t_b+t_w)$ for ${\bf P}_-$. Hence, the non-zero phase factor $\phi$ allows us to open different gaps for the two Fermi points. In particular, we choose 
\bq
t_b\ll t_w, 
\eq
so the two Fermi points have a large energy difference, as shown in Fig. \ref{fig:EnergyGaps}. By restricting to low enough energy scales, of the order of $\Delta E_+$, the dynamics of ${\bf P}_-$ will be frozen and it can be neglected. To demonstrate this consider the ground state, $\ket{\text{gs}}$, of the system and two excited states, $\ket{e_+}$ corresponding to the lowest energy excitation at ${\bf P}_+$ and $\ket{e_-}$ corresponding to ${\bf P}_-$. Next, we assign the energy gaps $\Delta E_+$ and $\Delta E_-$ between each of the excited states and the ground state. Assume that the system is initially prepared in the ground state $\ket{\text{gs}}$. Consider a small perturbation in the system that couples the ground state to both excited states with equal strength $\Omega$ of the order of $\Delta E_+$. This perturbation has as an effect a negligible population to be transferred to $\ket{e_-}$ and most of the dynamics to take place only between  $\ket{\text{gs}}$ and $\ket{e_+}$. Indeed, by adiabatic elimination we find that the maximum population of $\ket{e_-}$ state at all times is of the order of $(\Omega/\Delta E_-)^2$, which we also verified numerically. Hence, we can safely neglect the ${\bf P}_-$ Fermi point as long as the perturbations acting on the system satisfy $\Omega \ll \Delta E_-$. 

Finally, the interaction $U$-term, $H_U$, of Hamiltonian (\ref{Ham1}) is local and acts as a repulsion between the fermions in the same unit cell. In the continuous approximation it takes the form
\be
H_U \approx U\int d^2r b^\dagger(\rr)b(\rr)w^\dagger(\rr)w(\rr),
\ee
where for $U\ll \Delta E_-$ we only consider fermionic modes around the ${\bf P}_+$ Fermi point. Combining all the components together we can write the continuum limit of (\ref{Ham1}), up to an overall energy shift, in the following way
\be
H\! \approx \! \int \!d^2 r \!\Big[\psi^\dagger\! \big(c\SIGMA\cdot \p + \sigma_z Mc^2\big)\psi +{g^2 \over 2} j^\mu j_\mu\Big],
\label{Ham2}
\ee
where $\psi(\rr) = (b(\rr) \,\, w(\rr))^T$ is the Dirac spinor, $\sigma_\mu$ are the Pauli operators with $\SIGMA=(\sigma_y,\sigma_x)$, $\p = (-i\partial_x,-i\partial_y) $, $j_x = \bar \psi \gamma_x\psi = \psi^\dagger \sigma_y \psi$, $j_y = \bar \psi \gamma_y\psi = \psi^\dagger \sigma_x \psi$, $\bar \psi =\psi^\dagger \sigma_z$ and $j_0 = \psi^\dagger \psi$. Moreover, $g^2={U\over 3}$, $c={3\over 2}t$ and $M = {2\over 3}{t_b \over t^2}$. Hence, the nearest neighbour tunnelling coupling corresponds to the speed of light, the next-to-nearest neighbour tunnelling coupling gives rise to the mass $M$ of the Dirac fermions and the lattice fermion interaction corresponds directly to the current-current interaction of the Thirring model.

{\em From Thirring model to Chern-Simons theory:--} Hamiltonian (\ref{Ham2}) exactly describes the massive Thirring model in $2+1$ dimensions. We now employ the path integral formalism to show the connection of this model to Chern-Simons-Maxwell theory \cite{Fradkin}. By applying a Wick rotation on the temporal coordinate, we can write the Euclidean partition function of the Thirring model given in \rf{Ham2} as
\be
Z_\text{Th}=\!\int \!\mathcal{D}\bar{\psi}\mathcal{D}\psi \exp\!\Big\{\!\!-\!
\!\!\int \!\! d^{3}x 
\Big[\bar{\psi}(c\;\displaystyle{\not}\partial -M c^{2})\psi-\frac{g^{2}}{2}
j^{\mu}j_{\mu}\Big]\Big\}.
\ee
We can introduce a vector field $a_{\mu}$ through the
following identity
\bq
&&\!\!\!\!\!\!\!\exp\left(\int d^{3}x \;
\frac{g^{2}}{2}\;j^{\mu}j_{\mu}\right) =  \\ 
&&\int
\mathcal{D}a_{\mu}\exp\left[-\int d^{3}x \;
\left(\frac{1}{2}a^{\mu}a_{\mu}+g\;
j^{\mu}a_{\mu}\right)\right],
\nonumber
\eq
so that the exponent of the partition function becomes quadratic with respect to the fermionic field. We can now integrate out the spinor fields
\bq
&&\!\!\!\!\!\!\!\int\mathcal{D}\bar{\psi}\;\mathcal{D}\psi \;\exp\left[- \int
c\; d^{3}x \; \bar{\psi}\left(\displaystyle{\not}\partial+\frac{g}{c}
\;\displaystyle{\not}a - M c\right)\psi\right]= \nonumber \\ 
&&
\exp\big\{ -S_\text{eff}[a]\big\},
\eq
and obtain an $a_\mu$-dependent effective action given by
\be
S_\text{eff}[a] = -\log \left[\det
\left(\displaystyle{\not}\partial+\frac{g}{c}\;\displaystyle{\not}a - M c\right)\right] .
\ee
Upon applying a Pauli-Villars regularisation \cite{Niemi,Redlich} to the effective action we obtain a parity violating term
\be
S_\text{eff}[a] = \frac{i g^{2}}{8 \pi c} \frac{M c}{|M c|}
\int d^{3}x\;\epsilon^{\lambda\mu\nu}a_{\lambda}
\partial_{\mu}a_{\nu}+ \mathcal{O}\left(\frac{\partial}{Mc}\right),
\label{eqn:finale}
\ee
which is the Abelian Chern-Simons action up to corrections of order ${\partial}/{Mc}$. As we are interested in the behaviour of the ground state of the system, which belongs to its low energy sector, the $\mathcal{O}\left({\partial}/{Mc}\right)$ and higher order terms will have a negligible contribution. Expression \rf{eqn:finale} comes from one-loop calculations of Feynman diagrams. However, the Coleman-Hill theorem \cite{Hill} guarantees that the Chern-Simons is the dominant term and it receives no further concreteness at higher loops. For convenience we take ${M c}/{|M c|}$ to be positive.

Next, we introduce an interpolating action $S_\text{I}[a, A]$ given by
\bq
&&S_\text{I}[a, A]=
\\ \nonumber
&&\int d^{3}x
\Big(\frac{1}{2}\;a^{\mu}a_{\mu}-i \epsilon^{\lambda\mu\nu}
a_{\lambda}
\partial_{\mu}A_{\nu}+\frac{2\pi i c}{g^{2}}\epsilon^{\lambda\mu\nu}
A_{\lambda} \partial_{\mu}A_{\nu}\Big),
\eq
where $A_{\mu}$ is an Abelian gauge field. By integrating the partition function of $S_\text{I}[a, A]$ with respect to $A_\mu$ or with respect to $a_\mu$ it is possible to prove \cite{Deser} the following equivalence between the two different partition functions
\bq
&&Z_\text{I}=\int\mathcal{D}a_{\mu}\mathcal{D}A_{\mu}e^{-S_\text{I}[a, A]}=
\nonumber \\
\nonumber 
&&\!\!\!\!\!\!\int\mathcal{D}a_{\mu}\exp\left[ - \int d^{3}x \;
\left(\frac{i g^{2}}{8
\pi c}\epsilon^{\lambda\mu\nu}a_{\lambda}
\partial_{\mu}a_{\nu} \!+\!\frac{1}{2}a^{\mu}a_{\mu}\right)\right]=\\ \nonumber 
&&\!\!\!\!\!\! \int\!\! \mathcal{D}A_{\mu} \!\exp \!\left[- \!\!\!\int \!\! d^{3}x 
\left(\frac{2\pi i \;c}{g^{2}}\epsilon^{\lambda\mu\nu}A_{\lambda}
\partial_{\mu}A_{\nu} \!+\! \frac{1}{4}F_{\mu\nu}F^{\mu\nu}\right)\right] =\\ 
&&Z_\text{CSM}.
\label{eqn:finalpartition}
\eq
Summing up, through the bosonisation mechanism, we have shown that the low energy sector of the $2+1$-dimensional Thirring model is equivalent to a Chern-Simons-Maxwell theory. In the standard spacetime with Lorentzian signature the topological action appears with the coupling ${2 \pi c}/{g^{2}}$. Hence, to enhance the Chern-Simons action over the Maxwell term we need to make the coupling $g^2$ small, but non-zero. Alternatively, if we are interested in obtaining the electromagnetic action in $2+1$ dimensions then we need to make the coupling $g^2$ large. Here we are interested in the case where the topological action is dominant. The Chern-Simons term makes the gauge theory massive, with a correlation length that decreases proportionally to $g^{2}$. In particular, the corresponding electric and magnetic fields die off exponentially fast away from the sources. Nevertheless, the field $A_{\nu}$ can take non-zero values everywhere much like the Aharonov-Bohm effect. Note that rescaling the integrated $A_\mu$ field in \rf{eqn:finalpartition} by powers of $g$ can make the pre factor of the Chern-Simons theory analytic in the limit $g\rightarrow 0$, while still the ratio of the couplings between the Chern-Simons and the Maxwell term, and thus our above analysis, would remain the same. 

{\em Measurement of topological order:--} Finally, we would like to identify the topological order of the tight-binding model given in \rf{Ham1}. Initially, we consider the Chern-Simons theory. The relevant physical observables should be operators that are gauge-invariant as well as metric independent. For that we take the Wilson loop operators
\be
W(L) = \exp \left({iq\over g}\oint_L A_\mu dx^\mu\right),
\label{eqn:WilsonLoop}
\ee
where $L$ is an arbitrary link in $2+1$ spacetime, possibly having many strands. Here, $q$ is the charge associated with the quasiparticle excitations of the Thirring model. It was shown in \cite{PolyakovGauss,Giavarini} that the expectation value of the Wilson loop, $\langle W(L) \rangle_\text{CSM}$, can be expressed in terms of the linking number $\Phi_L$, known also as the Gauss integral, of the link $L$ as $\langle W(L) \rangle_\text{CSM} = \exp \big(\pm iq^2{\Phi_L / (8\pi)}\big)$. For two loops that are linked once, it is $\Phi_L=1$. Then \rf{eqn:WilsonLoop} corresponds to the braiding of two quasiparticles with fractional statistical angle $\theta_\mathrm{stat} =q^2/8\pi$ \cite{Fradkin}. For a single unknotted loop $L_0$ it is $\Phi_{L_0} =0$, so the expectation value becomes equal to $\langle W(L_0) \rangle_\text{CSM} = 1$. If $L_0$ lies completely on the spatial surface of the Chern-Simons theory, i.e. having no time component, then this expectation value is evaluated with respect to the ground state of the system $\ket{\Psi_\text{CSM} }$ and it gives
\be
\bra{\Psi_\text{CSM} } W(L_0) \ket{\Psi_\text{CSM} } = 1.
\label{eqn:loop}
\ee
Hence, the Chern-Simons theory has a ground state that is stabilised in terms of Wilson  operators of all possible loop shapes. For non-trivial ground states or loop operators this condition can be satisfied only by states that are condensates of all possible loops. Such loop condensate states are topologically ordered as they exhibit non-zero topological entropy \cite{Levin,PachosBook} and they have non-trivial topological degeneracy when the system is wrapped around the torus \cite{Freedman}. These two characteristics are the main identification tools of topological order. 

Condition \rf{eqn:loop} allows us to determine if the tight-binding model with a low energy behaviour described by the Thirring model is topologically ordered or not. It was shown by Fradkin and Schaposnik \cite{Fradkin} that the expectation value of the Wilson loop can be expressed in terms of fermionic observables of the Thirring model, i.e.
\be
\langle W(L) \rangle_\text{CSM} = \langle \exp \left(iq\int_\Sigma dS_\mu \bar\psi \gamma^\mu\psi\right)\rangle_\text{Th},
\label{eqn:expectation}
\ee
where $\Sigma$ is a surface bounded by the loop $L$. We can employ this connection to express condition \rf{eqn:loop} of topological order in terms of fermionic observables. Consider a spatial surface, $\Sigma_0$, of the Thirring model. The flux of the fermionic current through $\Sigma_0$ is given in terms of the current $j_0 = \psi^\dagger \psi$ as
\be
\int_{\Sigma_0} dS_\mu \bar\psi \gamma^\mu\psi = \int_{\Sigma_0} dS\left[b(\rr)^\dagger b(\rr)+w(\rr)^\dagger w(\rr)\right].
\label{eqn:current}
\ee
In terms of the tight-binding model the flux of the current $j_0$ through $\Sigma_0$ becomes the sum of the fermionic densities of both species at the sites enclosed by $\Sigma_0$. Hence, the expectation value of the exponential of these populations with respect to the ground state of the tight-binding model, $\ket{\Psi_\text{TB}}$, is given by
\be
\bra{\Psi_\text{TB}}\exp\Big[i q \sum_{{\bf i}\in\Sigma_0}\big(b_{\bf i}^\dagger b^{}_{\bf i} + w_{\bf i}^\dagger w^{}_{\bf i}\big)\Big]\ket{\Psi_\text{TB}} = 1.
\label{eqn:cond}
\ee
In other words $\ket{\Psi_\text{TB}}$ is a superposition of states that are eigenstates of the enclosed population operators with eigenvalues that are multiples of $2\pi/q$. Hence, a measurement of the population can reveal the value of $q$, which is yet theoretically undetermined~\cite{Essler}. 
\begin{figure}
\includegraphics[width=0.35\textwidth]{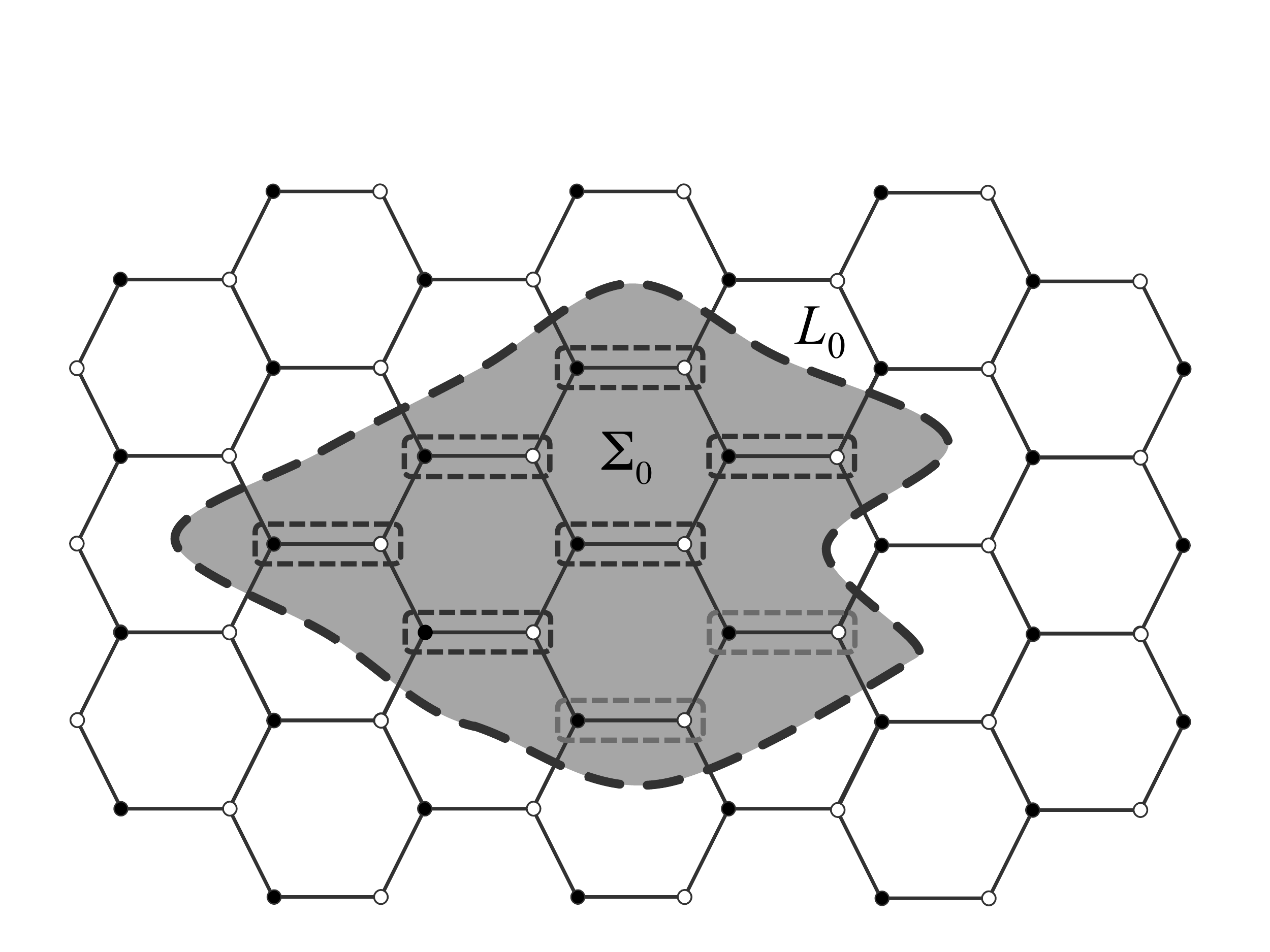}
\caption{\label{fig:measurement} A part of the honeycomb lattice with a loop $L_0$ depicted that encloses the surface $\Sigma_0$. The loop encloses a number of lattice unit cells, of which the populations $b_{\bf i}^\dagger b^{}_{\bf i}$ and $w_{\bf i}^\dagger w^{}_{\bf i}$ are measured. When the expectation value of $\exp[i q\sum_{{\bf i} \in \Sigma_0}(b_{\bf i}^\dagger b^{}_{\bf i}+w_{\bf i}^\dagger w^{}_{\bf i})]$ tends to a non-zero value for arbitrarily large areas $\Sigma_0$ then the ground state of the system is necessarily topologically ordered.}
\end{figure}
One can now directly determine if the tight-binding model is topologically ordered. In Fig. \ref{fig:measurement} we depict an area $\Sigma_0$ of the lattice bounded by a loop $L_0$. Care has been taken so that $L_0$ does not cut cells in half as they are considered as a single point in space during the continuous approximation. Then \rf{eqn:cond} corresponds to measuring the populations of $b$ and $w$ fermions, $b_{\bf i}^\dagger b^{}_{\bf i}$ and $w_{\bf i}^\dagger w^{}_{\bf i}$, in all sites within the region $\Sigma_0$ of the tight-binding model, constructing their sum and then averaging their exponential over different realisations of the lattice experiment. Note that if the coupling $g^2$ is large and the Maxwell term is dominant over the Chern-Simons action then $\langle W(L_0)\rangle \approx e^{-\kappa |\Sigma_0|}$, where $\kappa$ is some positive constant and $|\Sigma_0|$ is the area enclosed by the loop $L_0$ \cite{Polyakov}. This quantity decreases exponentially fast as the area of the loop is increased due to the large quantum fluctuations in the enclosed fermionic populations. This area law behaviour reveals the charge confinement of $2+1$-dimensional QED \cite{Wilson} and it can be directly demonstrated with our scheme.

{\em Conclusions:--} In this letter we have presented a tight-binding model that gives rise, in the low energy limit, to Abelian Chern-Simons theories. We extended a version of Haldane's model, with imbalanced masses of the resulting Dirac fermions so that one of them is adiabatically eliminated. In this limit interactions between fermions exactly reproduce the Thirring model. Upon bosonisation the latter model is equivalent to the Abelian Chern-Simons theory. A direct method to measure the topological order of the system is proposed that requires local density measurements of the fermions of the tight-binding model. These measurements can determine the invariance of the ground state under applications of arbitrary Wilson loop operators of the model resulting from bosonisation. The study of the quasiparticle excitations of this model as well as its generalisation to non-Abelian Chern-Simons theories \cite{FradkinNonAb} is a fundamental problem with practical applications to topological quantum technologies \cite{PachosBook}.

A possible experimental realisation of the tight-binding model can be given in terms of spin-dependent potentials, in the same lines as Refs. \cite{Alba2,Alba,Goldman}. There interspecies tunnelling along the honeycomb lattice is activated by Raman assisted tunnelling, which can imprint complex phase factors as the ones we require here \cite{Alba2}. The interactions between fermions are restricted only within the unit cell, and thus need to be independent of the tunnelling couplings. For that one can employ optically induced $p$-wave Feshbach resonance to manipulate the collisional couplings $U$ \cite{Pachos,PachosChern}. Alternatively, if two out-of-phase spin-dependent potentials are employed to trap the atomic states $b$ and $w$, independently, then one can bring the $b$ and $w$ atoms of the same cell in arbitrarily close proximity, thus improving the tunability of their interaction. Finally, the local atom density measurements necessary to identify the topological order can be performed with well established techniques \cite{Bloch}. Relation \rf{eqn:cond} can then be verified for arbitrary surfaces $\Sigma_0$ with geometric characteristics that are large compared to the correlation length of the system \cite{EmilioGezaJaun}. 

Note that an alternative approach to obtain fractional quantum Hall physics by introducing interactions in the Haldane model has been recently presented in \cite{Mudry}, though that model is analytically intractable.

{\em Acknowledgements:--} JKP would like to thank Gunnar Moller for inspiring conversations. This work was supported by EPSRC.

\end{document}